\documentclass[letterpaper]{article} 
\usepackage{aaai25}  
\usepackage{times}  
\usepackage{helvet}  
\usepackage{courier}  
\usepackage[hyphens]{url}  
\usepackage{graphicx} 
\usepackage{natbib}  
\usepackage{caption} 
\usepackage{algorithm}
\usepackage{algorithmic}
\usepackage{subfig}
\usepackage{listings}
\usepackage{xcolor}
\usepackage{booktabs} 

\definecolor{codegreen}{rgb}{0,0.6,0}
\definecolor{codegray}{rgb}{0.5,0.5,0.5}
\definecolor{codepurple}{rgb}{0.58,0,0.82}
\definecolor{backcolour}{rgb}{0.95,0.95,0.92}

\lstdefinestyle{jsonstyle}{
    backgroundcolor=\color{backcolour},
    basicstyle=\scriptsize\ttfamily, 
    commentstyle=\color{codegreen},
    keywordstyle=\color{magenta},
    numberstyle=\tiny\color{codegray},
    stringstyle=\color{codepurple},
    numbers=none, 
    breakatwhitespace=false,
    breaklines=true,
    captionpos=b,
    frame=single,
    keepspaces=true
}

\usepackage{newfloat}
\usepackage{listings}
\DeclareCaptionStyle{ruled}{labelfont=normalfont,labelsep=colon,strut=off} 
\floatstyle{ruled}
\newfloat{listing}{tb}{lst}{}
\floatname{listing}{Listing}
\title{YT-30M: A multi-lingual multi-category dataset of YouTube comments}
\author{
   Hridoy Sankar Dutta
}
\affiliations{
    School of Information Technology, Deakin University\\

    hridoy.dutta@deakin.edu.au
}

\begin{document}
\nocopyright
\maketitle

\begin{abstract}
This paper introduces two large-scale multilingual comment datasets, YT-30M (and YT-100K) from YouTube. The analysis in this paper is performed on a smaller sample (YT-100K) of YT-30M. Both the datasets: YT-30M (full) and YT-100K (randomly selected 100K sample from YT-30M) are publicly released for further research. YT-30M (YT-100K) contains $32236173$ ($108694$) comments posted by YouTube channel that belong to YouTube categories. Each comment is associated with a video ID, comment ID, commentor name, commentor channel ID, comment text, upvotes, original channel ID and category of the YouTube channel (e.g., ‘News \& Politics’, ‘Science \& Technology’, etc.).
\end{abstract}

%
\begin{links}
    \link{Datasets}{https://huggingface.co/datasets/hridaydutta123/YT-100K}
\end{links}

\section{Introduction}

The recent popularity of video-sharing platforms such as YouTube has revolutionized how people consume and create content in the online world. With a massive number of monthly active users and a significant increase in engagement, YouTube plays a critical role in digital marketing and content consumption \cite{rieder2023making}.

The creation of a multilingual dataset for YouTube is important for understanding cultural nuances and sentiment expressions that vary from one language to another. Adding the multicategory feature is additionally important for comprehending the reasons behind these nuances and expressions across different types of content. Unlike a few social media platforms such as Twitter \cite{chang2023roeoverturned, pfeffer2023just,shaik2023sentiment,comito2023multimodal} and Facebook \cite{aljabri2023machine, perrotta2021behaviours,ernala2020well}, which have been thoroughly studied by the academic community in previous years, today YouTube accounts for a significant share of this market, as it is the second most visited website globally after Google.

This paper presents the first large-scale multilingual multi-category dataset, YT-30M\footnote{Only preliminary analysis on YT-100K dataset is provided in this version of the paper. The analysis and use cases can be easily performed on the main dataset with high computational power.} for comment classification tasks on YouTube. To the best of our knowledge, YT-30M is the largest publicly available YouTube dataset for academic research. Table \ref{table_sample} shows sample multilingual comments from our collected dataset for 5 different languages. YT-30M contains comments from more than 50 different languages. Figure \ref{figure_json} displays a sample YouTube comment from our dataset along with all the features collected for that comment. 

\begin{table}[!ht]
\centering
\begin{tabular}{ll}
\toprule
Language & Text\\
\midrule
English & You girls are so awesome!! \\
French & Ceci est un commentaire \\
German & Das Geräusch eurer Klingel erinnert mich an ei... \\
Spanish & jajajaj esto tiene que ser una brom \\
Portuguese & nossa senhora!!!!!!!!!!!!!!!!!!!!!!!!!!!!!!!!!...\\
\bottomrule
\end{tabular}
\caption{Sample multilingual comments posted on YouTube in 5 different languages.}
\label{table_sample}
\end{table}

\begin{figure}
\begin{lstlisting}[style=jsonstyle]
{
    "videoID": "ab9fe84e2b2406efba4c23385ef9312a",
    "commentID": "488b24557cf81ed56e75bab6cbf76fa9",
    "commentorName": "b654822a96eae771cbac945e49e43cbd",
    "commentorChannelID": "2f1364f249626b3ca514966e3ef3aead",
    "comment": "ich fand den Handelwecker am besten",
    "votes": 2,
    "originalChannelID": "oc_2f1364f249626b3ca514966e3ef3aead",
    "category": "entertainment"
}
\end{lstlisting}
\caption{A YouTube comment from YT-30M}
\label{figure_json}
\end{figure}

\begin{figure*}[!ht]
    \centering
    \includegraphics[width=\linewidth]{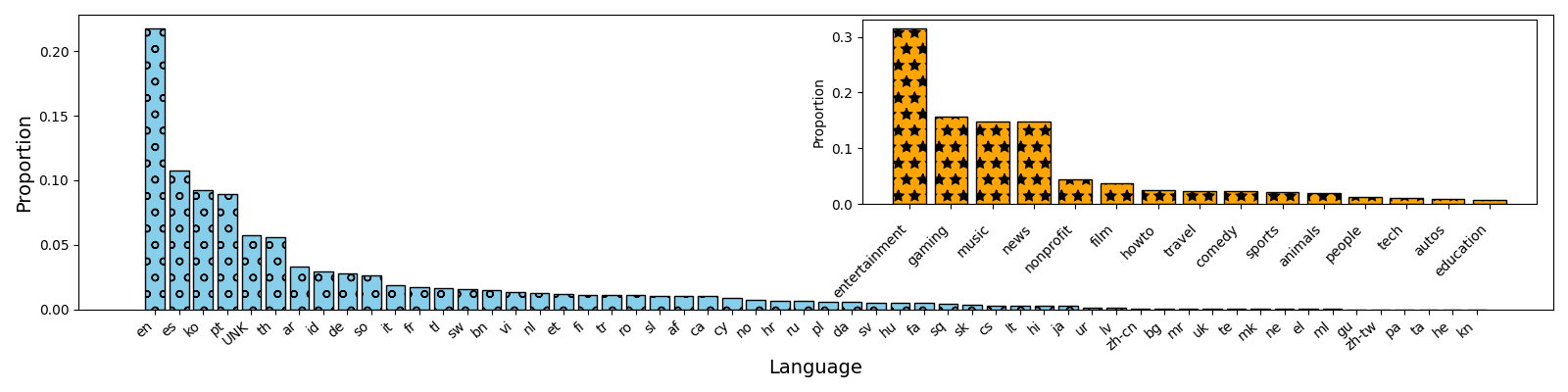}
    \caption{The main plot shows the proportion of languages detected in YouTube comments. The inset plot shows the proportion of YouTube categories.}
    \label{fig:language}
\end{figure*}
\begin{figure*}[!ht]
    \centering
    \includegraphics[width=\linewidth]{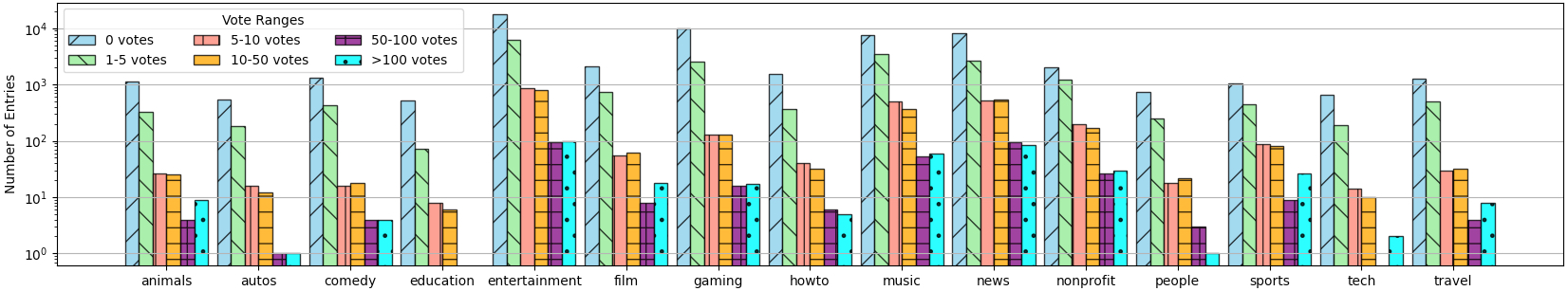}
    \caption{Upvotes distribution for YouTube categories.}
    \label{fig:upvotes}
\end{figure*}

\section{Dataset characteristics}
Each entry in the dataset is related to one comment for a specific YouTube video in the related category with the following columns: videoID, commentID, commentorName, 
 commentorChannelID, comment, votes, originalChannelID, category. Each field is explained below:
 \begin{enumerate}
     \item \textbf{videoID:} represents the video ID in YouTube.
     \item \textbf{commentID:} represents the comment ID.
     \item \textbf{commentorName:} represents the name of the commentor.
     \item \textbf{commentorChannelID:} represents the ID of the commentor.
     \item \textbf{comment:} represents the comment text.
     \item \textbf{votes:} represents the upvotes received by that commment.
     \item \textbf{originalChannelID:} represents the original channel ID who posted the video.
     \item \textbf{category:} represents the category of the YouTube video.
 \end{enumerate}
Table \ref{tab:youtube_comments_statistics} details the statistics of the YT-30M dataset. We have 32,236,173 (32 million) unique YouTube comments from 178,027 videos posted by 20,568,637 (20 million) unique commentors. It is important to note that all Personally Identifiable Information (PII) has been redacted in the released dataset.
\begin{table}[h]
    \centering
    \caption{Statistics of YT-30M}
    \begin{tabular}{@{}lcccccc@{}}
        \toprule
        Field & YT-30M \\ \midrule
        Number of videos & 178027  \\
        Number of comments & 32236173  \\
        Number of commentors  & 20568637\\
        \bottomrule
    \end{tabular}
    \label{tab:youtube_comments_statistics}
\end{table}
\section{Data analysis}
\begin{figure}[!htb]
    \centering
    \subfloat[Sentiment distribution of YouTube channel category.]{%
        \includegraphics[width=0.5\textwidth]{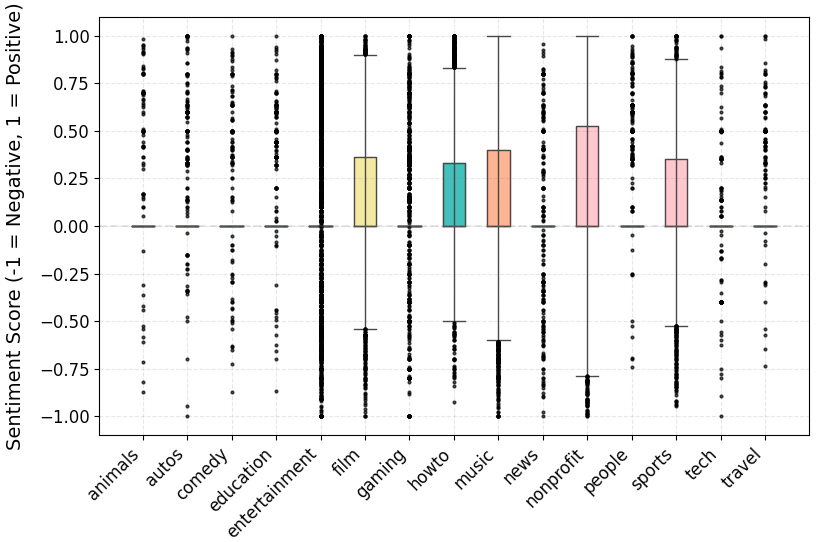} 
    }
    \hfill 
    \subfloat[Comment length distribution of YouTube channel category.]{%
        \includegraphics[width=0.5\textwidth]{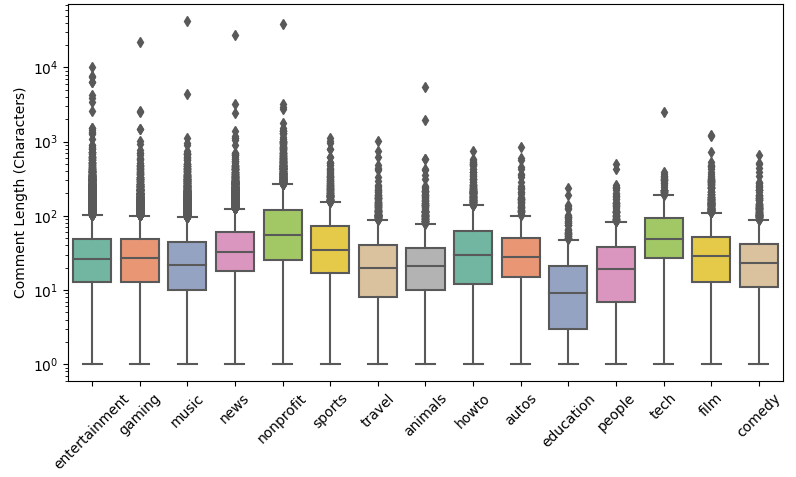} 
    }
    \hfill 
    \subfloat[Wordcloud of comment text.]{%
        \includegraphics[width=0.5\textwidth]{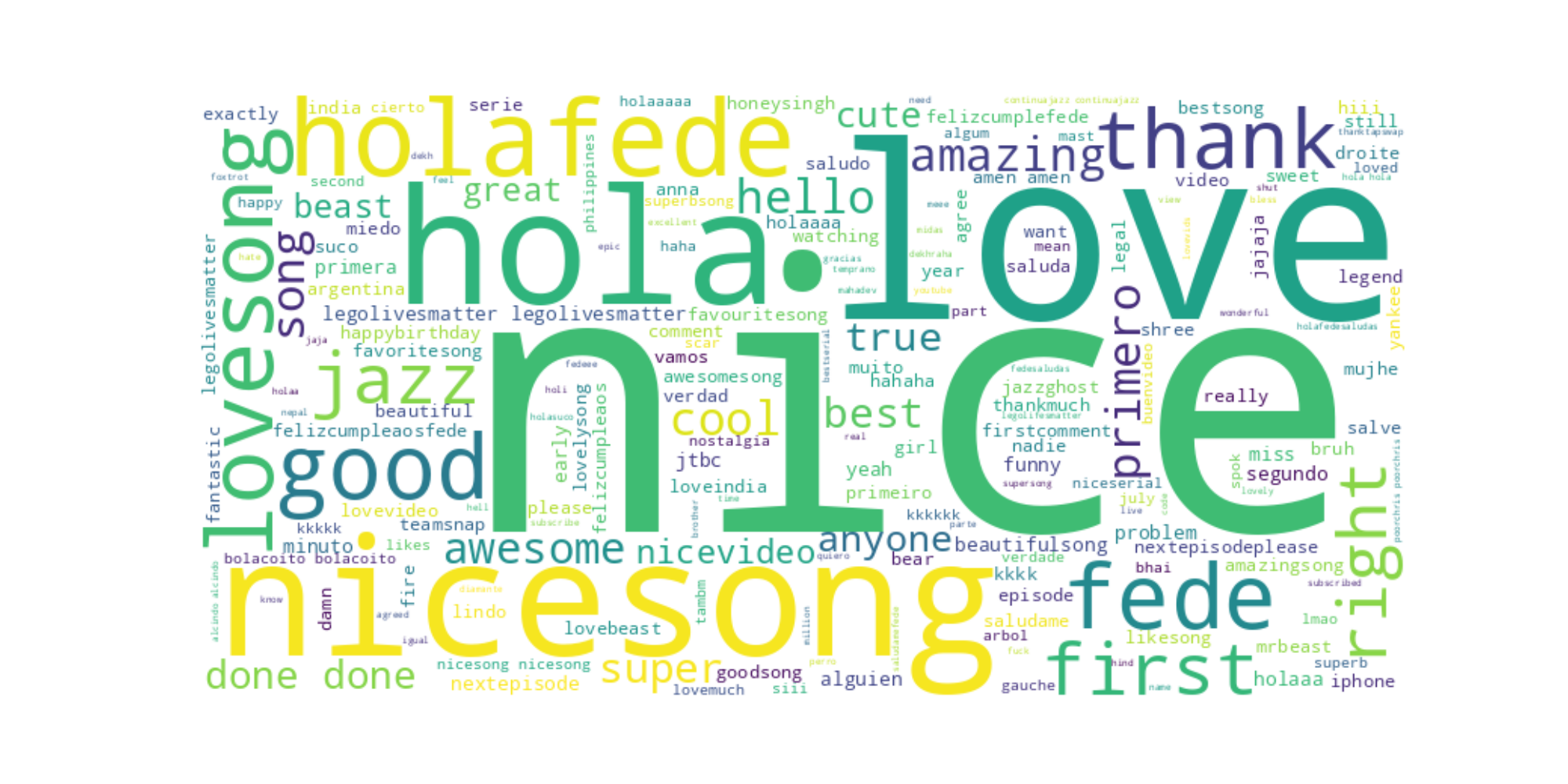} 
    }
    \caption{Analysis of our collected dataset.}
    \label{fig:images}
\end{figure}

Due to computational limitations, we performed our analysis on the small dataset YT-100K. YT-100K contains a random selection of 100,000 comments from the YT-30M dataset. Figure \ref{fig:language} shows the proportion of different languages detected in our comment dataset, while Figure \ref{fig:language} (inset) illustrates the distribution of categories across the dataset. Our multilingual dataset includes comments from more than 50 languages, which were detected using the Python langdetect library. Each comment is mapped to a category corresponding to its associated YouTube channel. The category assigned to each comment helps in understanding broader societal trends present within each category. Every comment in the dataset includes the number of upvotes it has received on YouTube. Figure \ref{fig:upvotes} displays the upvote distribution for YT-30M across different YouTube categories. We observe that certain categories, such as music, news, and people, which are more engaging to habitual users (subscribers/members), receive a higher number of upvotes (more than 100 votes) compared to other categories. Additionally, we notice a steep decay in the number of upvotes for each subsequent vote range (5-10, 10-50, etc.), which may be attributed to comments not meeting engagement criteria or simply going unnoticed due to the large volume of comments submitted to a video.
Further analysis is presented in Figure \ref{fig:images}. In Figure \ref{fig:images}(a), we show the sentiment distribution of the comments in our dataset. The sentiment score ranges from -1 (negative) to +1 (positive), with 0 being neutral. We found that categories such as ``education'' and ``how-to'' have many comments clustered near zero, indicating less emotional engagement. In contrast, ``news'' appears to be the category with the broadest range of sentiment scores. Figure \ref{fig:images}(b) illustrates the comment length distribution from our dataset. We observe that although most comments are very short, categories like ``news'' and ``nonprofit'' tend to have longer comments compared to others. This could be due to these categories typically having more discussions than categories such as ``gaming'', which often contain more reactive comments. A wordcloud depicting common words in our comment dataset is shown in Figure \ref{fig:images}(c).

\textbf{Data anonymity.} This work collects only publicly available YouTube comments, and all PII (personally identifiable information) are redacted for anonymity. 

\section{Release and Contributions}
The YT-100K dataset is available on the Hugging Face platform\footnote{\url{https://huggingface.co/datasets/hridaydutta123/YT-100K}}. The YT-30M dataset can be obtained by requesting the author of this dataset. The author encourage researchers working in the domain of Natural Language Processing and Social Network Analysis to perform various interesting analyses and modeling on this dataset.

\bibliography{CameraReady/LaTeX/aaai25}

\end{document}